\documentstyle[12pt]{article}

\begin{document}
\title{Cosmic Tests for a More Explicit Equivalence Principle}

\author{Rafael A. Vera\thanks{email: rvera@buho.dpi.udec.cl}\\
Deptartamento de Fisica\\
Fac. de C.Fis y Mat. Universidad de Concepcion\\
Casilla 4009. Concepcion. Chile}

\maketitle

\begin{abstract}
According to this principle, the relativistic changes occurring to the
bodies, after velocity changes, cannot be detected by observers moving
with them because bodies and stationary radiations change in
identical proportion after identical circumstances, i.e, because bodies
and stationary radiations have identical relativistic laws with respect
to any fixed observer. Effectively the theoretical properties of
particle models made up of stationary radiations agree with special
relativity, quantum mechanics and the gravitational (G) tests. They fix
lineal properties for all of them: the G fields, the black holes (BHs)
and the universe. The BHs, after absorbing radiation, must return to
the gas state. An eventual universe expansion cannot change any
relative distance because the G expansion of matter occurs in identical
proportion. This fixes a new kind of universe. In it matter evolves in
closed cycles, between gas and BH states and vice versa, indefinitely.
Galaxies and clusters must evolve rather cyclically between luminous
and black states. Most of the G potential energy of a matter cycle must
be released around neutron star and black hole boundaries. Nuclear
stripping reactions would transform G energy into nuclear and kinetic
energies. This accounts for many non well explained phenomena in
astrophysics. This work has been published, in more detail, in a book.
\end{abstract}

\section {The New Universe Fixed by the Explicit Equivalence
Principle}

It is obvious that to study the universe we must use the most reliable
physical principles. So far the Equivalence Principle is the best
tested principle in physics and, therefore, this is the most solid base
for any theory, provided that it does not depend on arbitrary
assumptions. The explicit form of this principle makes
possible to avoid the dependence on any arbitrary assumption because
everything is fixed by properties of quantums of radiation. In this
way everything in the universe is fixed by this principle that
ultimately depends rather well known radiation properties. Viceversa,
the good fit of the real universe with the theoretical one turns
out to be a good test either for the EEP or for the radiation
properties.

The underlying physical theory based on the EEP has been treated more 
extensively, and in more detail, in previous works published and 
presented in several meetings during the last 23 years~\cite{v1}${-
}$~\cite{v12}. This one has been briefly summarized in another 
contribution presented in the MG8 meeting~\cite{v12}. The present work 
is a review of the work done to test the new astrophysical and 
cosmological contexts fixed by the EEP~\cite{v11}. 

In previous works it has been proven that, if all of the intergalactic
distances were increasing with the time, in the same proportion, the
G potential and length of the particle model would also increase with
the time, in identical proportion as that of the universe. Then it
would be not possible to find 
a strictly invariable standard body that does not expands itself in 
identical proportion as  in the universe. This leads to conclude that {\it
An eventual universe expansion cannot change the relative values of all
of them : distances, velocities, temperatures and cosmological red shifts.
This means that the physical laws are also invariable after universe
expansion}. This result turns out to be a self-consistency test of the EEP.

Consequently, according to the EEP, {\it the Hubble red shift cannot
depend on the time and, therefore, the age and lifetime of the universe
are not limited by an eventual universe expansion}.

On the other hand, according to the linear field equation fixed by the 
EEP, the new angular momentum law fixes the critical escape angles of 
the photons and relativistic particles going away from a neutron star 
with $GM/r >1/2$, called {\it black hole} (BH). Such angles are very 
small but not null.  Thus the light emission rate of a massive BH is 
negligible compared with its absorption rate. So the average relativistic
mass-energy of each of its nucleons must increase with the time, after
the radiation absorption. Thus, sooner or later the BH must become
unstable and explode. Such explosion must transform the relativistic
mass-energy of its nucleons into the potential and kinetic energies of the
hydrogen rich cloud generated after such expansion. The last one would be
condensed at faster rates over the dead stars and planetesimals
traveling around the BH. This would produce {\it star clusters} that,
sooner or later, after radiation emission, must return to BH states,
and so on indefinitely~\cite{v2}${\!,\,}$~\cite{v4}${\!,\,}$~\cite{v11}. 

A chain of BH explosions must produce a {\it luminous galaxy} that, 
sooner or later, must end as a {\it black galaxy} (BG). The last one
would also have global properties resembling a BH.

Similar mechanism holds for a chain of luminous galaxy formations
which would produce {\it clusters}.

In general, macro-systems should evolve between rather luminous 
and black states, and vice versa, indefinitely.  In this way {\it the 
average universe entropy would remain constant}. 

\section{Cosmic Tests for the Explicit Equivalence 
Principle}

The new kind of universe provides crucial tests for the EEP. In 
particular:

{\bf Test Ia}.- In order that the radiation lost by all of the luminous
galaxies can be equal to the one absorbed by the black ones, the average
number of the last ones must be of higher order of magnitude than the
first ones. Thus most of the universe should be in more dense and dark
states.

{\bf Test Ib}.- The same result comes out from the higher average
density of the universe derived from the constants $G$ and $H$.
Then most of the universe mass must be in the state of black
galaxy (BG). Observe that only the higher G fields of massive BGs and
their associations can account for the higher speeds observed in
luminous galaxies. This would solve one of the cluster missing mass
problem in astrophysics. Notice that in the partially black galaxies,
the BHs and dead star remnants, produced after 
star evolution, can also account for their (apparent) missing mass 
problems.

{\bf Test II}.- All of the stages of the matter evolution cycles must
be present in the sky.

The new galaxies generated after chains of BH explosions, must have 
the highest fractions of random angular momentum and maximum 
volumes of low density stars recently formed with hydrogen not 
contaminated with metals. They correspond with the {\it red elliptical 
galaxies}. By canceling random angular momentum and loosing energy, 
they must become denser and bluer, with an increasing number of dead 
stars in their borders. They must show disc regions of higher {\it net} 
angular momentum densities that, for this reason, don't contract as fast 
as the random ones of the spherical bulges. Thus {\it the disc and spiral 
galaxies} correspond to intermediate stages of galaxy contraction.

The final luminous bulge of a galaxy, surrounded by dead star remnants, 
must be in lower G potentials. Thence it must emit light with higher G 
redshifts\footnote {The existence of an additional red shift occurring 
during the light trip through strong field gradients of star remnants 
cannot be ruled out}. Due to the smaller volume of the luminous region, 
and to its lower absolute luminosity, the relative luminosity changes 
caused by star's explosions are higher. Due to the high average masses 
and densities of the final luminous  stars, they must show more 
explosive events and emit higher proportions of ultraviolet light 
compared with a normal galaxy. Those stages correspond with the {\it 
AGNs and quasars}. 

Then it may be concluded that most of the quasar's red shift would be 
intrinsic (not cosmological) one, and that their absolute luminosity's 
would be very small compared with ordinary galaxies. They would 
normally correspond to the last luminous periods of  galaxy cycles. 

{\bf Test III}.- According to point I, most of {\it the low temperature
cosmic radiation background} should come from BGs that are cooled down
by BHs and the rest of the universe. It has been found that most of
such photons should come from BGs located in long distance ranges of
the order of magnitude of the Hubble Radius~\cite{v11}. This is
consistent both with the low value of such temperature and with {\it
its high isotropy}.

{\bf Test IV}.-  The net G energy yield  per unit of mass-
energy~\cite{v7}${\!,\,}$~\cite{v11} captured by a central mass $M$ 
turns out to be equal to $1-exp[-GM/r]$. From this relation it is
inferred that :
 
a)  {\it The net G energy yield in a matter cycle is of a higher order of 
magnitude than the one of nuclear fusion. 

b)  b)Most of it must be released in the strongest G fields of neutron 
stars (NSs) and BHs, either steadily or explosively}. 

Since the G binding energies per unit of mass of the neutrons in a NS
are much higher than the nuclear binding energies of ordinary nuclei,
then in principle {\it neutron stripping reactions} (NSRs) should occur
after the impact of atomic nuclei on a NS (or BH). The NS would capture
neutrons and reject protons or proton rich nuclei. Globally, NSRs would
transform G potential energy into kinetic and nuclear potential
energies, mainly in the states of relativistic protons of high magnetic
rigidities. This kind of reaction would convert He and heavier atoms
into renewed H, at the cost of the lower final G potential of some
neutrons captured by the NS. This mechanism is consistent with 
the energies and composition of {\it cosmic rays}. They are obviously 
much richer both in protons and proton rich nuclei. 

The main fields of a NS (or a BH) would drive the external plasma 
towards its polar regions where it would fall rather vertically. After 
NSRs, according to momentum conservation, the rejection angles of the 
relativistic particles are even smaller than the incident ones. Thus
they must have higher probabilities to escape from the G field 
along the magnetic axis. These particle beams correspond with the {\it 
cosmic jets} observed in central regions of galaxies. Such regions are 
just those with the highest probabilities for the existence of both:
rather naked BHs and gas clouds. Such jets, in their paths, can excite
fluorescence on gas and particles. 

In the rather naked BHs that are left over after some supernovas, the 
eventual precession of the magnetic poles can drive the jets along
cones producing visible rings in their intersections with spherical
shells of matter ejected from earlier explosions~\cite{v11}. These
jets are obvious in the Hubble Space Telescope pictures of:
SNA1987A, the Hourglass, the Egg Nebulas, and in PRC95-24 (C.
Burrows, J. Hester, J. Morse and NASA). They have been
available in internet, from NASA.

{\bf Test V}.-  From points I and IV), the most powerful stars should
have a NS inside of them\footnote {The probable mechanisms of NS
formation in stars have been discussed in the reference $2$.}.
Their energies should come both from nuclear fusion and neutron
stripping. The last reactions must convert large fractions of the G
work into nuclear potential energy that would be used up in the
external shell~\cite{v7}${\!,\,}$~\cite{v11}. This would prevent all
of them: local overheating, neutrino cooling and star
collapse (urca process).
 
{\it The new star model} has a better defined mass-luminosity 
relation~\cite{v7}${\!,\,}$~\cite{v11}, which is roughly proportional
to $M^{3.7}$. This one is consistent with that of {\it main sequence 
stars}. This is also consistent with the higher densities and
temperatures of such stars, and with {\it the low neutrino luminosity
of the Sun and its magnetic activity}.

\section{Conclusions}

The EEP fixes more simple and unified physical and astrophysical 
contexts, starting from a single quantum and ending with the universe.
They are self-consistent and consistent with the observed facts.
The high unity observed this theory comes from the fact that
everything turns out to be fixed by common properties of radiations.
This way also reduces the number of independent variables and the
uncertainties due either to arbitrary assumptions or experimental
errors (or limitations). 

The use of a well defined reference frame, which is equivalent to 
generalize the language used in special relativity, prevents the current 
errors due to the ambiguity of the ordinary language used in current 
literature. 

I started this work about 24 years ago from very simple but fundamental 
questions, trying just to understand by myself the basic phenomena in
nature. Little by little the few first ideas have grown up into an
extremely simple and self-consistent theory that can be used as a guide
for understanding, in a unified way, a wide range of phenomena occurring
in the wonderful universe in which we live~\cite{v11}.

Curiously, the truth may be much more simple than we expect. This may
be just the reason for which we normally do not see it so easily.

\end{document}